# QUANTUM MEMORY FOR PHOTONS IN CASE OF MANY CLOSE LYING EXCITON RESONANCES IN SOLIDS.

A. D. Gazazyan[1,a], E. A. Gazazyan[1] and A. G. Margaryan[2]

[1] Institute for physical Research, NAS of Armenia, Ashtarak-2 0203, Armenia
[2] Yerevan State Medical University after Mkhitar Heratsi , Koryun 2 St., Yerevan, 0025, Armenia
[a] e-mail: Alfred@ipr.sci.am

**Abstract.** The possibility of storage of quantum information with photons is studied in the case of resonant transitions via many close lying exciton levels in a solid with impurity $\Lambda$-atoms. The upper levels of the impurity atom form resonant Fano states, similar to the autoionization atomic states, due to the configuration interaction with the continuum of the exciton band. In this case slowing of light pulses is shown to be realistic, in the presence of the control field, down to the group velocity much lower than that in vacuum. The possibility of storage and reconstruction of a quantum pulse is studied in the case of the instantaneous switching on/off of the control field. It is shown that the signal quantum pulse cannot be stored undistorted for differing values of Fano parameters and for non-zero two-photon detuning and decay rate between the lower levels (decoherence). However, for small difference of the Fano parameters and for small values of the two-photon detuning and the decoherence there is no distortion in the case where the length of the pulse is much longer than the linear absorption (amplification) length, so the shape and quantum state of the light pulse can be restored.

1. ## INTRODUCTION

Quantum memory for photons and the storage and reconstruction of quantum information represents one of the topical problems of modern science. In recent works devoted to this problem, the propagation, storage, and reconstruction of a light pulse are considered in a medium with $\Lambda$ atoms that exhibit discrete levels. The control field generates coherence between the lower levels in the $\Lambda$ system, which is maintained in the medium after the field is switched off. Owing to the electromagnetically induced transparency (EIT) [1, 2] of the medium, the information on the input signal is stored in the medium after the control field is switched off and is restored upon the control field's subsequently being switched on as an output pulse whose shape is identical to the shape of the input pulse due to the coherence between the lower states [3–8].

The propagation of a quantum pulse in a medium with $\Lambda$ atoms is considered in [3–6] in the adiabatic limit of the atomic interaction with the fields. It is demonstrated that the shape and quantum state of the reconstructed pulse are identical to the shape and state of the original pulse. The reconstructed pulse propagates at the group velocity, which is significantly lower than the velocity of light in free space. New quantum states (dark and bright polaritons) that represent the superposition of the quantum electromagnetic field and a collective atomic component are introduced in [3–6]. Using this elegant approach, the authors study the propagation of a quasi-particle (dark polariton) that is transformed into a quantum field or a spin wave in the medium in various limiting cases and is controlled with external laser radiation.

The possibility for the storage and reconstruction of classical and quantum pulses is considered in [9, 10] in the case of an instantaneous nonadiabatic switching on/off of the fields in a medium with $\Lambda$ atoms. It is demonstrated that, under certain conditions, the storage and reconstruction of a pulse whose shape is identical to the shape of the original pulse is possible upon the nonadiabatic atomic interaction with the fields. The polariton states and their properties are also considered in [9] in the case of the nonadiabatic switching of the fields.

From the practical point of view, it is expedient to study the propagation, storage, and reconstruction of light pulses in solids. The EIT theory for solids is developed in [11] where homogeneous and inhomogeneous broadening of the line in one- and two-photon transitions plays a significant role. A decrease in the group velocity of a propagating light pulse is demonstrated. The behavior of the dark polaritons, which are introduced in [3–6], in the solid and the possibility for the storage of the quantum data under the conditions of inhomogeneous broadening are studied in [12].

Aiming at the development of a UV laser, Kuznetsova et al. [13] studied the possibility for the suppression of the absorption in the excited state, in the presence of an additional external field, based on the EIT phenomenon [2] in laser crystals. A high-lying discrete level or the conduction bands in the crystal are considered as the excited state.

However, highly excited states of the impurity atom can be located inside an exciton band of the crystal. In this case, the configuration interaction of these states with the continuum of the exciton band gives rise to resonant states with a relatively large width that are similar to the autoionization Fano states in atoms [14]. The formation of the exciton resonance leads to asymmetry in the absorption spectrum of the impurity atom in the crystal [15–18].

In previous paper [19] the possibility was shown for the storage and reconstruction of quantum information in the case of one isolated exciton resonance with assumption that the Fano parameters are equal, while the two-photon detuning and decay rate between the lower levels (decoherence) are zero. It was demonstrated that the shape and quantum state of the light pulse are preserved in the course of its propagation and reconstruction. In this work, we study the possibility for the storage of quantum memory for photons in the case of many close lying exciton resonances in a solid. It will be shown that for differing values of Fano parameters and for none zero two photon detuning and decay rate between the lower levels (decoherence) the quantum pulse is distorted. However, for slightly differing Fano parameters and for small two photon detuning and small decoherence, in the case where the length of the pulse is much larger than the linear absorption (amplification) length, there is no distortion and the shape and quantum state of the light pulse can be restored.

We also consider the reconstruction of quantum pulse in the presence of a control field in a crystal with impurity $\Lambda$ atoms, when the upper state is located in the exciton band. Then, the configuration interaction of this state with the continuum of the exciton band leads to the formation of the resonant Fano state similar to the autoionization atomic state [14]. A similar state of the exciton resonance can also be formed under the action of the laser (control) field on the low-lying state, as in the case of the laser induced continuum structure (LICS) [20] in atoms.

Section 2 is devoted to the formation of exciton resonances in a solid. The Hamiltonian of the problem is written and the system of Heisenberg–Langevin equations for the collective atomic operators, as well as the equation of propagation of a quantum pulse in the presence of exciton resonances, is derived.

In Section 3 we study the possibility for the storage and reconstruction of quantum information using the exciton resonance upon the nonadiabatic interaction of the fields. It is demonstrated that the induced coherence between the lower discrete states of the system is maintained after the control field is switched off. When the control field is switched on, this coherence restores the quantum pulse, whose shape and quantum state are identical to those of the input pulse.

Section 4 is devoted to the simplest case of two close lying isolated exciton resonances. Finally, we summarize and discuss the results obtained.

## 2. FORMATION OF EXCITON RESONANCES IN SOLIDS AND BASIC EQUATIONS.

We consider a crystalline medium with N impurity Λ-atoms in the case when the upper states of the impurity Λ-atoms are located in the exciton band. The upper levels of the impurity atom form a Fano resonant state similar to the autoionization atomic state due to the configuration interaction with the continuum of the exciton band. By analogy with the autoionization resonances [14], the wave function of the exciton states can be represented as

$$\Psi_\lambda = \left[\sum_\nu \frac{\tan \delta_\nu}{\pi \widetilde{U}^*_{\lambda\nu}} \Phi_{\nu\lambda} - \psi_\lambda\right] \cos \delta , \qquad (1)$$

where $\Phi_{\nu\lambda}$ is the modified discrete state given by

$$\Phi_{\nu\lambda} = \widetilde{\psi}_\nu + P \int \frac{dE}{\lambda - E} \widetilde{U}^*_{\nu E} \psi_E . \qquad (2)$$

Here the symbol "P" denotes the principal value of the integral, $U_{\nu,E} = \langle \psi_E | U | \psi_\nu \rangle$ is the matrix element of the configuration interaction U, $\psi_E$ is the wave function of the smooth continuum (exciton band),

$$\widetilde{\psi}_\nu = \sum_k S_{k\nu} \psi_k , \qquad (3)$$

$$\widetilde{U} = S^+ U , \qquad (4)$$

$$\tan \delta_\nu = -\frac{\pi |\widetilde{U}_{\nu\lambda}|^2}{\lambda - \widetilde{E}_\nu}, \qquad (5)$$

and $\delta_\nu$ is the phase shift due to interaction with the continuum corresponding to the ν resonance, while δ is the total phase shift:

$$\tan \delta = \sum_\nu \tan \delta_\nu = -\frac{\pi}{z(\lambda)}, \qquad (6)$$

with

$$z(\lambda) = \left[\sum_\nu \frac{|\widetilde{U}_{\nu\lambda}|^2}{\lambda - \widetilde{E}_\nu}\right]^{-1} . \qquad (7)$$

Here S is a unitary matrix. Elements of this matrix and $\widetilde{E}_\nu$ can be obtained by soling the system

$$E_k S_{\nu\lambda} + \sum_{k'} \Delta_{kk'}(\lambda) S_{k'\nu} = S_{k\nu} \widetilde{E}_\nu , \qquad (8)$$

where

$$\Delta_{kk'}(\lambda) = P \int dE \frac{U_{kE} U^+_{Ek'}}{\lambda - E} . \qquad (9)$$

To study the possibility of quantum memory for photons and of the storage and subsequent reconstruction of quantum information, we consider the propagation of a quantum light pulse at the carrier frequency ω in a crystal consisting of N impurity Λ-atoms in the presence of a coherent control field at the carrier frequency $\omega_L$. The Λ-atoms have two discrete lower states 1 and 2 and the high-lying states located in the exciton band of the crystal (Fig. 1). The configuration interaction U of the upper discrete states with the continuum of the exciton band generates a resonant Fano state like the autoionization atomic state [14]. The coherent control field with the interaction constant Ω provides the coupling of the exciton resonant state $\Psi_\lambda$ with the discrete level 1, while the quantum field couples the exciton resonant state $\Psi_\lambda$ with the discrete level 2 (Fig. 1). The exciton resonance is formed when a real discrete state $\{\psi_k\}$ is mixed with the continuum of the exciton band due to the configuration interaction U or owing to the LICS [20].

The positive-frequency part $\widehat{E}^+(z_j,t)$ of the operator of quantum pulse for the electromagnetic field propagating along the *z* axis is written as

$$\widehat{E}^+(z_j,t) = -i\left(\frac{2\pi\hbar\omega}{V}\right)^{1/2} \hat{a}_j(t) e^{i(kz_j - \omega t)} , \qquad (10)$$

where $\hat{a}_j(t)$ are slowly varying operators and the subscript j determines the position of an atom along the *z* axis; V is the quantization volume.

In the interaction representation, by neglecting the damping, the Hamiltonian of the system of Fig. 1 takes the form

$$H = \hbar \sum_j \int d\lambda \{\hat{\sigma}^j_{\lambda,1}(t)\Omega^j_\lambda(t)e^{i(\lambda-\varepsilon_1-\omega_L)t} + \hat{\sigma}^j_{\lambda,2}(t)g_\lambda \hat{a}_j(t)e^{i(\lambda-\varepsilon_2-\omega)t} + \text{h.c.}\}, \quad (11)$$

where the summation with respect to j involves all the atoms on the *z* axis, and $\hat{\sigma}^j_{i,k}$ are atomic operators given by

$$\hat{\sigma}^j_{i,k} = |i\rangle^j \langle k|^j. \quad (12)$$

Here $\Omega^j_\lambda(t)$ is the matrix element of the transition of the j the atom from the state 1 to the state of exciton resonance $\Psi_\lambda$ induced by the control coherent electromagnetic field and

$$g_\lambda = \left(\frac{2\pi\hbar\omega}{V}\right)^{1/2} \hat{d}_{2\lambda}, \quad (13)$$

Here $\hat{d}_{2\lambda}$ is the matrix element of the operator of the dipole moment for the transition from the state 2 to the state of exciton resonance $\Psi_\lambda$. To describe the evolution of the atomic operators, we obtain, based on the Hamiltonian (7), the following system of Heisenberg–Langevin equations:

$$\frac{d\hat{\sigma}^j_{1,1}(t)}{dt} = -i \int d\lambda \left\{ i\frac{\gamma^j_1}{2}\hat{\sigma}^j_{\lambda,\lambda}(t) + \hat{\sigma}^j_{\lambda,1}(t)\Omega^j_\lambda(t) - \Omega^{j*}_\lambda(t)\hat{\sigma}^j_{1,\lambda}(t) \right\} + \hat{F}^j_1(t) \quad (14a)$$

$$\frac{d\hat{\sigma}^j_{2,2}(t)}{dt} = -i \int d\lambda \left\{ i\frac{\gamma^j_2}{2}\hat{\sigma}^j_{\lambda,\lambda}(t) + \hat{\sigma}^j_{\lambda,2}(t)g_\lambda \hat{a}_j(t) - g_\lambda \hat{a}^+_j(t)\hat{\sigma}^j_{2,\lambda}(t) \right\} + \hat{F}^j_2(t) \quad (14b)$$

$$\frac{d\hat{\sigma}^j_{\lambda,\lambda'}(t)}{dt} = i\left\{ \left(\lambda - \lambda' + i\frac{\gamma^j_{\lambda,\lambda'}}{2}\right)\hat{\sigma}^j_{\lambda,\lambda}(t) + \hat{\sigma}^j_{\lambda',1}(t)\Omega^j_\lambda(t) + \hat{\sigma}^j_{\lambda,2}(t)g_{\lambda'}\hat{a}_j(t) - \Omega^{j*}_\lambda(t)\hat{\sigma}^j_{1,\lambda'}(t) - \right.$$
$g_{\lambda'}*\alpha j + t\sigma_{2,\lambda'}jt + F_{\lambda\lambda'}j(t)$ (14c)

$$\frac{d\hat{\sigma}^j_{1,\lambda}(t)}{dt} = -i\left\{ \left(\lambda - \varepsilon_1 - \omega_L - i\frac{\gamma^j_{1,\lambda}}{2}\right) \hat{\sigma}^j_{1,\lambda}(t) - \Omega^j_\lambda(t) \hat{\sigma}^j_{1,1}(t) - \hat{\sigma}^j_{1,2}(t) g_\lambda \hat{a}_j(t) + \right.$$
$d\lambda'\Omega\lambda'jt\ \sigma\lambda,\lambda'jt + F1,\lambda jt$ (14d)

$$\frac{d\hat{\sigma}^j_{2,\lambda}(t)}{dt} = -i\left\{ \left(\lambda - \varepsilon_2 - \omega - i\frac{\gamma^j_{2,\lambda}}{2}\right) \hat{\sigma}^j_{2,\lambda}(t) - \Omega^j_\lambda(t) \hat{\sigma}^j_{2,1}(t) - \hat{\sigma}^j_{2,2}(t) g_\lambda \hat{a}_j(t) + \right.$$
$d\lambda'\ \sigma\lambda,\lambda'jt\ g\lambda\ \alpha jt + F\ 2,\lambda jt$ (14e)

$$\frac{d\hat{\sigma}^j_{2,1}(t)}{dt} = i\left\{ \left(\nu + i\frac{\gamma^j_c}{2}\right) \hat{\sigma}^j_{2,1}(t) + \int d\lambda \left( \hat{\sigma}^j_{2,\lambda}(t)\Omega^{j*}_\lambda(t) - \hat{\sigma}^j_{\lambda,1}(t)g_\lambda \hat{a}_j(t) \right) \right\} + \hat{F}^j_{2,1}(t) \quad (14f)$$

Here $\gamma_i$ and $\gamma_{i,k}$ are the longitudinal and transverse decay constants, respectively, $\hat{F}^j_i(t)$ and $\hat{F}^j_{i,k}(t)$ are δ-correlated quantum operators of the Langevin noise, and

$$\nu = \varepsilon_2 + \omega - \varepsilon_1 - \omega_L \quad (15)$$

is the two-photon detuning of resonance.

To solve the problem of the propagation of a quantum pulse, we assume that the interaction with atoms $g_\lambda \hat{a}_j(t)$ is significantly weaker than the interaction of the control field $\Omega^j_\lambda(t)$ and that the photon density in the input pulse is substantially less than the atomic density [5]. In this case, atomic equations (14) can be analyzed using the perturbation theory with respect to $g_\lambda \hat{a}_j(t)$. In the zero order, we may assume that $\langle\hat{\sigma}_{11}\rangle = \langle\hat{\sigma}_{\lambda\lambda'}\rangle = 0$, and $\langle\hat{\sigma}_{22}\rangle = 1$. With allowance for the fact that $\Omega_\lambda^j(t), \gamma_i^j, \gamma_{i,k}^j$, and $\gamma_c^j$ weakly depend on j, in the first order with respect to $g_\lambda \hat{a}_j(t)$ we obtain the following expression based on Eq. (14f):

$$\frac{d \hat{\sigma}_{2,1}^j(t)}{dt} = i\left\{\left(\nu + i\frac{\gamma_c}{2}\right) \hat{\sigma}_{2,1}^j(t) + \int d\lambda\, \hat{\sigma}_{2,\lambda}^j(t)\Omega_\lambda^*(t)\right\} + \hat{F}_{2,1}^j(t) \qquad (16)$$

On the other hand, from Eq. (14e) we have

$$\hat{\sigma}_{2,\lambda}^j(t) = \hat{\sigma}_{2,1}^j(t)\frac{\Omega_\lambda(t)}{\lambda - \varepsilon_2 - \omega - i\frac{\gamma_{2\lambda}}{2}} + \frac{g_\lambda}{\lambda - \varepsilon_2 - \omega - i\frac{\gamma_{2\lambda}}{2}}\hat{a}_j(t) - i\frac{\hat{F}_{2,\lambda}^j(t)}{\lambda - \varepsilon_2 - \omega - i\frac{\gamma_{2\lambda}}{2}}. \qquad (17)$$

Assuming that the amplitudes are slowly varying on the interval $\Delta z$ containing $N_z \gg 1$ atoms, we can introduce continuous atomic variables [5]:

$$\hat{\sigma}_{\mu,\nu}(z,t) = \frac{1}{N_z}\sum_j \hat{\sigma}_{\mu,\nu}^j(t) \qquad (18)$$

With allowance for expression (18), expressions (16) and (17) are represented as

$$\frac{d\hat{\sigma}_{2,1}(z,t)}{dt} = i\left\{\left(\nu + i\frac{\gamma_c}{2}\right) \hat{\sigma}_{2,1}(z,t) + \int d\lambda\, \hat{\sigma}_{2,\lambda}(z,t)\Omega_\lambda^*(t)\right\} + \hat{F}_{2,1}(z,t), \qquad (16')$$

$$\hat{\sigma}_{2,\lambda}(z,t) = \hat{\sigma}_{2,1}(z,t)\frac{\Omega_\lambda(t)}{\lambda - \varepsilon_2 - \omega - i\frac{\gamma_{2\lambda}}{2}} + \frac{g_\lambda}{\lambda - \varepsilon_2 - \omega - i\frac{\gamma_{2\lambda}}{2}}\hat{a}(z,t) - i\frac{\hat{F}_{2,\lambda}(z,t)}{\lambda - \varepsilon_2 - \omega - i\frac{\gamma_{2\lambda}}{2}}. \qquad (17')$$

The evolution of the Heisenberg operator of quantum field $\hat{a}(z,t)$ can be described using the propagation equation in the approximation of slowly varying amplitude:

$$\left(\frac{\partial}{\partial t} + c\frac{\partial}{\partial z}\right)\hat{a}(z,t) = iN\int g_\lambda \hat{\sigma}_{2\lambda}(z,t)d\lambda \qquad (19)$$

Substituting the expression for $\hat{\sigma}_{2,\lambda}(z,t)$ from (17') into (16') and (19), we obtain:

$$\frac{\partial \hat{\sigma}_{21}(z,t)}{\partial t} = i\hat{\sigma}_{21}(z,t)\left(\nu + i\frac{\gamma_c}{2} + \int d\lambda\frac{|\Omega_\lambda(t)|^2}{\lambda - \varepsilon_2 - \omega - i\frac{\gamma_{2\lambda}}{2}}\right) + i\left(\int d\lambda\frac{\Omega_\lambda^*(t)g_\lambda}{\lambda - \varepsilon_2 - \omega - i\frac{\gamma_{2\lambda}}{2}}\right)\hat{a}_j(t) + \mathscr{F}_{21}(z,t) \qquad (20)$$

$$\left(\frac{\partial}{\partial t} + c\frac{\partial}{\partial z}\right)\hat{a}(z,t) = iN\hat{a}(z,t)\int d\lambda\frac{g_\lambda^2}{\lambda - \varepsilon_2 - \omega - i\frac{\gamma_{2\lambda}}{2}} + iN\hat{\sigma}_{21}(z,t)\int d\lambda\frac{g_\lambda\Omega_\lambda(t)}{\lambda - \varepsilon_2 - \omega - i\frac{\gamma_{2\lambda}}{2}} + \mathscr{F}_\alpha(z,t) \qquad (21)$$

where

$$\mathscr{F}_{21}(z,t) = \int d\lambda\,\frac{\Omega_\lambda^*(t)}{\lambda - \varepsilon_2 - \omega - i\frac{\gamma_{2\lambda}}{2}}\hat{F}_{2,\lambda}(z,t) + \hat{F}_{2,1}(z,t) \qquad (22)$$

$$\mathscr{F}_\alpha(z,t) = N\int d\lambda\,\frac{g_\lambda}{\lambda - \varepsilon_2 - \omega - i\frac{\gamma_{2\lambda}}{2}}\hat{F}_{2,\lambda}(z,t) \qquad (23)$$

It is demonstrated in [14] that the ratios of the probabilities of transitions to the continuum state with structure $|\Omega_\lambda(t)|^2$ and $g_\lambda^2 |\langle\hat{a}_j(t)\rangle|^2$ to the probabilities of transitions to the state of smooth continuum $|\tilde{\Omega}_\lambda(t)|^2$ and $\tilde{g}_\lambda^2 |\langle\hat{a}_j(t)\rangle|^2$ are

$$\frac{|\Omega_\lambda(t)|^2}{|\tilde{\Omega}_\lambda(t)|^2} = \frac{\left[\prod_\nu(\lambda - \tilde{E}_\nu) + \frac{1}{2}\sum_\nu q_\nu^L \tilde{\Gamma}_\nu \prod_{k\neq\nu}(\lambda - \tilde{E}_\nu)\right]^2}{\prod_\nu(\lambda - s_\nu^*)(\lambda - s_\nu)} \qquad (24a)$$

$$\frac{|g_\lambda(t)|^2}{|\tilde{g}_\lambda(t)|^2} = \frac{\left[\prod_\nu(\lambda - \tilde{E}_\nu) + \frac{1}{2}\sum_\nu q_\nu \tilde{\Gamma}_\nu \prod_{k\neq\nu}(\lambda - \tilde{E}_\nu)\right]^2}{\prod_\nu(\lambda - s_\nu^*)(\lambda - s_\nu)}, \qquad (24b)$$

where the quantities $s_\nu$ are the roots of the polynomial:

$$\prod_\nu(\lambda - \tilde{E}_\nu) - \frac{i}{2}\sum_\nu \tilde{\Gamma}_\nu \prod_{k\neq\nu}(\lambda - \tilde{E}_k) = 0 \quad (\text{Im}\,s_\nu > 0), \qquad (25)$$

Here $\nu$ and k correspond to exciton resonances and $q_L$ and q are the corresponding parameters of the Fano asymmetry represented as:

$$q_\nu^L = \frac{\langle\phi_{\nu,\lambda}|\hat{\Omega}_\lambda(t)|\psi_1\rangle}{\pi u_\lambda^* \langle\psi_\lambda|\hat{\Omega}_\lambda(t)|\psi_1\rangle} \qquad (26a)$$

$$q_\nu = \frac{\langle\phi_{\nu,\lambda}|\hat{g}_\lambda|\psi_2\rangle}{\pi u_\lambda^* \langle\psi_\lambda|\hat{g}_\lambda|\psi_2\rangle} \qquad (26b)$$

Expressions (24a) and (24b) describe the Beutler–Fano profile for the photoabsorption cross-sections in the vicinity of the exciton resonance for various values of the Fano parameter.

At large Fano parameters, this profile exhibits a developed maximum. It follows from expressions (24a) and (24b) that

$$\Omega_\lambda(t) = \widetilde{\Omega}_\lambda(t) \frac{\prod_\nu (\lambda - \widetilde{E}_\nu) + \frac{1}{2}\sum_\nu q_\nu^L \widetilde{\Gamma}_\nu \prod_{k\neq\nu}(\lambda - \widetilde{E}_\nu)}{\sqrt{\prod_\nu (\lambda - S_\nu^*)(\lambda - S_\nu)}} \qquad (27a)$$

$$g_\lambda(t) = \widetilde{g}_\lambda(t) \frac{\prod_\nu (\lambda - \widetilde{E}_\nu) + \frac{1}{2}\sum_\nu q_\nu \widetilde{\Gamma}_\nu \prod_{k\neq\nu}(\lambda - \widetilde{E}_\nu)}{\sqrt{\prod_\nu (\lambda - S_\nu^*)(\lambda - S_\nu)}} , \qquad (27b)$$

It is seen from these expressions that the photoabsorption matrix elements become equal to zero at the energies $\lambda = \lambda_j$ ($\lambda_j = 1, \ldots n$), where the quantities $\lambda_j$ are the roots of the polynomial:

$$\prod_\nu (\lambda - \widetilde{E}_\nu) + \frac{1}{2}\sum_\nu q_\nu^L \widetilde{\Gamma}_\nu \prod_{k\neq\nu}(\lambda - \widetilde{E}_k) = 0, \qquad (28a)$$

or

$$\prod_\nu (\lambda - \widetilde{E}_\nu) + \frac{1}{2}\sum_\nu q_\nu \widetilde{\Gamma}_\nu \prod_{k\neq\nu}(\lambda - \widetilde{E}_k) = 0, \qquad (28b)$$

("Fano windows"). This is related to the destructive interference of the transitions to the continuous spectrum [14].

Below we neglect the slow dependence of $\Omega$, $\Delta$, and $\Gamma$ on $\lambda$ ($\gamma_{2\lambda} \equiv \overline{\gamma}_2$). Substituting expressions (27a) and (27b), in equations (20) and (21), integrating with respect to $\lambda$ in the vicinity of the exciton resonance at large q, we obtain the following equations for the atomic operator of the coherence between lower levels $\hat{\sigma}_{21}(z,t)$ and for propagation of the quantum pulse $\hat{\alpha}(z,t)$:

$$\frac{\partial \hat{\sigma}_{21}(z,t)}{\partial t} = i\hat{\sigma}_{21}(z,t)\left(\nu + i\frac{\gamma_c}{2} + i|\widetilde{\Omega}(t)|^2 \beta_1^L(\omega)\right) - \widetilde{\Omega}^*(t)\widetilde{g}\beta_2(\omega)\hat{\alpha}(z,t) + \mathcal{F}_{2,1}(z,t) \qquad (20')$$

$$\left(\frac{\partial}{\partial t} + c\frac{\partial}{\partial z}\right)\hat{\alpha}(z,t) = -N\widetilde{g}^2\beta_1(\omega)\hat{\alpha}(z,t) - N\widetilde{g}\widetilde{\Omega}(t)\beta_2(\omega)\hat{\sigma}_{21}(z,t) + \mathcal{F}_\alpha(z,t) , \qquad (21')$$

where

$$\beta_1(\omega) = \pi\left(1 - i\sum_\nu \frac{1}{\text{Im}S_\nu} \frac{[\prod_k(S_\nu^* - \widetilde{E}_k) + \frac{1}{2}\sum_j q_j\widetilde{\Gamma}_j \prod_{k\neq\nu}(S_\nu^* - \widetilde{E}_k)]^2}{[\text{Re}S_\nu - \varepsilon_2 - \omega - i(\frac{\overline{\gamma}_2}{2} + \text{Im}S_\nu)]\prod_{k\neq\nu}(S_\nu^* - S_k)(S_\nu^* - S_k^*)}\right) \qquad (29a)$$

$$\beta_1^L(\omega) = \pi\left(1 - i\sum_\nu \frac{1}{\text{Im}S_\nu} \frac{[\prod_k(S_\nu^* - \widetilde{E}_k) + \frac{1}{2}\sum_j q_j^L\widetilde{\Gamma}_j \prod_{k\neq\nu}(S_\nu^* - \widetilde{E}_k)]^2}{[\text{Re}S_\nu - \varepsilon_2 - \omega - i(\frac{\overline{\gamma}_2}{2} + \text{Im}S_\nu)]\prod_{k\neq\nu}(S_\nu^* - S_k)(S_\nu^* - S_k^*)}\right) \qquad (29b)$$

$$\beta_2(\omega) = \pi\left(1 - i\sum_\nu \frac{1}{\text{Im}S_\nu} \frac{[\prod_k(S_\nu^* - \widetilde{E}_k) + \frac{1}{2}\sum_j q_j^L\widetilde{\Gamma}_j \prod_{k\neq\nu}(S_\nu^* - \widetilde{E}_k)][\prod_k(S_\nu^* - \widetilde{E}_k) + \frac{1}{2}\sum_j q_j\widetilde{\Gamma}_j \prod_{k\neq\nu}(S_\nu^* - \widetilde{E}_k)]}{[\text{Re}S_\nu - \varepsilon_2 - \omega - i(\frac{\overline{\gamma}_2}{2} + \text{Im}S_\nu)]\prod_{k\neq\nu}(S_\nu^* - S_k)(S_\nu^* - S_k^*)}\right) \qquad (29c)$$

and

$$\mathcal{F}_{2,1}(z,t) = \widetilde{\Omega}^*(t) \int \frac{d\lambda}{\lambda - \varepsilon_2 - \omega - i\frac{\overline{\gamma}_2}{2}} \frac{\prod_\nu (\lambda - \widetilde{E}_\nu) + \frac{1}{2}\sum_\nu q_\nu^L \widetilde{\Gamma}_\nu \prod_{k\neq\nu}(\lambda - \widetilde{E}_k)}{\sqrt{\prod_\nu (\lambda - S_\nu^*)(\lambda - S_\nu)}} \hat{F}_{2,\lambda}(z,t) + \hat{F}_{2,1}(z,t) \qquad (30a)$$

$$\mathcal{F}_\alpha(z,t) = N\widetilde{g} \int \frac{d\lambda}{\lambda - \varepsilon_2 - \omega - i\frac{\overline{\gamma}_2}{2}} \frac{\prod_\nu (\lambda - \widetilde{E}_\nu) + \frac{1}{2}\sum_\nu q_\nu \widetilde{\Gamma}_\nu \prod_{k\neq\nu}(\lambda - \widetilde{E}_k)}{\sqrt{\prod_\nu (\lambda - S_\nu^*)(\lambda - S_\nu)}} \hat{F}_{2,\lambda}(z,t) \qquad (30b)$$

Equations (20') and (21') are the basic equations for the problem of the propagation of electromagnetic radiation in crystals with impurity $\Lambda$-atoms in the case when the upper state is located in the exciton band.

Consider the possibility of storage and reconstruction of quantum information upon the instantaneous on–off switching of the fields. For this purpose we analyze the propagation of a quantum pulse in the medium in the case when the interaction of the control field $\Omega$ does not depend on the spatial coordinate z and does not change with time. Using the Fourier transform with respect to the spatial coordinate z, we obtain the following system of differential equations for the k-th mode of the quantum pulse $\hat{\alpha}(z,t)$ and atomic operator $\hat{\sigma}_{21}(z,t)$ based on Eqs. (20') and (21'):

$$\frac{d\hat{\sigma}_{21}(k,t)}{dt} = i\left(\nu + i\frac{\gamma_c}{2} + i|\widetilde{\Omega}(t)|^2 \beta_1^L(\omega)\right)\hat{\sigma}_{21}(k,t) - \widetilde{\Omega}^*(t)\tilde{g}\beta_2(\omega)\hat{\alpha}(k,t) + \mathscr{F}_{2,1}(k,t) \quad (31)$$

$$\frac{d\hat{\alpha}(k,t)}{dt} + ikc\hat{\alpha}(k,t) = -N\tilde{g}^2\beta_1(\omega)\hat{\alpha}(k,t) - N\tilde{g}\widetilde{\Omega}(t)\beta_2(\omega)\hat{\sigma}_{21}(k,t) + \mathscr{F}_{\alpha}(k,t) \quad (32)$$

### 3. STORAGE OF QUANTUM MEMORY FOR PHOTON UPON THE NONADIABATIC INTERACTION OF THE FIELDS.

We solve the system of differential equations (31) and (32) for the instantaneous switching-on of the fields at t=0 using the Laplace transform. After the reconstruction of the original functions, we obtain the following expressions for the Fourier components of the quantum field $\hat{\alpha}(z,t)$ and atomic operator $\hat{\sigma}_{21}(z,t)$ in terms of two waves with frequencies $\omega_{\pm}(k)$ respectively:

$$\hat{\alpha}(k,t) = \hat{\alpha}_+(k,t) + \hat{\alpha}_-(k,t), \quad (33)$$

$\hat{\alpha}_+(k,t) =$
$\frac{1}{2}\left(1 + \frac{k-\tilde{k}_0}{\sqrt{(k-k_1)(k-k_2)}}\right)e^{-i\omega_+(k)t}\hat{\alpha}(k,0) -$
$\frac{i}{c}\beta_2(\omega)\frac{N\tilde{g}\widetilde{\Omega}}{\sqrt{(k-k_1)(k-k_2)}}e^{-i\omega_+(k)t}\hat{\sigma}_{21}(k,0) + \int_0^t \mathbb{K}_1^{(+)}(k,t')e^{-i\omega_+(k)(t-t')}dt' \quad (34a)$

$\hat{\alpha}_-(k,t) =$
$\frac{1}{2}\left(1 - \frac{k-\tilde{k}_0}{\sqrt{(k-k_1)(k-k_2)}}\right)e^{-i\omega_-(k)t}\hat{\alpha}(k,0) +$
$\frac{i}{c}\beta_2(\omega)\frac{N\tilde{g}\widetilde{\Omega}}{\sqrt{(k-k_1)(k-k_2)}}e^{-i\omega_-(k)t}\hat{\sigma}_{21}(k,0) + \int_0^t \mathbb{K}_1^{(-)}(k,t')e^{-i\omega_-(k)(t-t')}dt' \quad (34b)$

and

$$\hat{\sigma}_{21}(k,t) = \hat{\sigma}_+(k,t) + \hat{\sigma}_-(k,t), \quad (35)$$

where

$\hat{\sigma}_+(k,t) = \frac{1}{2}\left(1 - \frac{k-\tilde{k}_0'}{\sqrt{(k-k_1)(k-k_2)}}\right)e^{-i\omega_+(k)t}\hat{\sigma}_{21}(k,0) - \frac{i}{c}\beta_2(\omega)\frac{\tilde{g}\widetilde{\Omega}^*}{\sqrt{(k-k_1)(k-k_2)}}e^{-i\omega_+(k)t}\hat{\alpha}(k,0) +$
$\int_0^t K_2^{(+)}(k,t')e^{-i\omega_+(k)(t-t')}dt' \quad (36a)$

$\hat{\sigma}_-(k,t) = \frac{1}{2}\left(1 + \frac{k-\tilde{k}_0'}{\sqrt{(k-k_1)(k-k_2)}}\right)e^{-i\omega_-(k)t}\hat{\sigma}_{21}(k,0) + \frac{i}{c}\beta_2(\omega)\frac{\tilde{g}\widetilde{\Omega}^*}{\sqrt{(k-k_1)(k-k_2)}}e^{-i\omega_-(k)t}\hat{\alpha}(k,0) +$
$\int_0^t K_2^{(-)}(k,t')e^{-i\omega_-(k)(t-t')}dt' \quad (36b)$

Here

$$\omega_{\pm}(k) = \frac{c}{2}\left(k - k_0 \pm \sqrt{(k-k_1)(k-k_2)}\right) \quad (37)$$

$$k_1 = -\frac{i}{c}\left[\beta_1^L(\omega)|\widetilde{\Omega}|^2 - \beta_1(\omega)N\tilde{g}^2 + 2i\beta_2(\omega)\sqrt{N}\tilde{g}|\widetilde{\Omega}| - i(\nu + i\frac{\gamma_c}{2})\right] \quad (38a)$$

$$k_2 = -\frac{i}{c}\left[\beta_1^L(\omega)|\widetilde{\Omega}|^2 - \beta_1(\omega)N\tilde{g}^2 - 2i\beta_2(\omega)\sqrt{N}\tilde{g}|\widetilde{\Omega}| - i(\nu + i\frac{\gamma_c}{2})\right] \quad (38b)$$

$$k_0 = \frac{i}{c}\left[\beta_1^L(\omega)|\widetilde{\Omega}|^2 + \beta_1(\omega)N\tilde{g}^2 - i(\nu + i\frac{\gamma_c}{2})\right] \quad (38c)$$

$$\tilde{k}_0 = -\frac{i}{c}\left[\beta_1^L(\omega)|\widetilde{\Omega}|^2 - \beta_1(\omega)N\tilde{g}^2 + i(\nu + i\frac{\gamma_c}{2})\right] \quad (38d)$$

$$\tilde{k}_0' = -\frac{i}{c}\left[\beta_1^L(\omega)|\widetilde{\Omega}|^2 - \beta_1(\omega)N\tilde{g}^2 - i(\nu + i\frac{\gamma_c}{2})\right] \quad (38e)$$

and
$$\mathbb{K}_1^{(+)}(k,t) = \frac{1}{2}\left(1 + \frac{k-\tilde{k}_0'}{\sqrt{(k-k_1)(k-k_2)}}\right)\mathcal{F}_\alpha(k,t) - \frac{i}{c}\frac{\beta_2(\omega)N\tilde{g}\tilde{\Omega}}{\sqrt{(k-k_1)(k-k_2)}}\mathcal{F}_{21}(k,t) \tag{39a}$$

$$\mathbb{K}_1^{(-)}(k,t) = \frac{1}{2}\left(1 - \frac{k-\tilde{k}_0'}{\sqrt{(k-k_1)(k-k_2)}}\right)\mathcal{F}_\alpha(k,t) - \frac{i}{c}\frac{\beta_2(\omega)N\tilde{g}\tilde{\Omega}}{\sqrt{(k-k_1)(k-k_2)}}\mathcal{F}_{21}(k,t) \tag{39b}$$

$$\mathbb{K}_2^{(+)}(k,t) = \frac{1}{2}\left(1 - \frac{k-\tilde{k}_0'}{\sqrt{(k-k_1)(k-k_2)}}\right)\mathcal{F}_{21}(k,t) - \frac{i}{c}\frac{\beta_2(\omega)\tilde{g}\tilde{\Omega}^*}{\sqrt{(k-k_1)(k-k_2)}}\mathcal{F}_\alpha(k,t) \tag{39c}$$

$$\mathbb{K}_2^{(-)}(k,t) = \frac{1}{2}\left(1 + \frac{k-\tilde{k}_0'}{\sqrt{(k-k_1)(k-k_2)}}\right)\mathcal{F}_{21}(k,t) + \frac{i}{c}\frac{\beta_2(\omega)\tilde{g}\tilde{\Omega}^*}{\sqrt{(k-k_1)(k-k_2)}}\mathcal{F}_\alpha(k,t) \tag{39d}$$

We can restrict consideration to small k, since the maximum value of the wave number $k_{max}$ is determined by the minimum length in the system [9]. For the storage of light in the medium, the pulse length in the medium must be shorter than the length of the medium. Then, $k_{max} \approx (v_g T)^{-1}$, where T is the pulse duration [9]. In the case of small k, expression (37) is simplified:

$$\omega_\pm(k) = \omega_\pm(0) + v_g^\pm k + i\chi^\pm k, \tag{40}$$

where

$$\omega_\pm(0) = \frac{c}{2}\left(-k_0 \pm \sqrt{k_1 k_2}\right) \tag{41a}$$

$$v_g^\pm = \frac{c}{2}\left(1 \mp \mathrm{Re}\frac{\tilde{k}_0'}{\sqrt{k_1 k_2}}\right) \tag{41b}$$

$$\chi^\pm = \mp \mathrm{Im}\frac{\tilde{k}_0'}{\sqrt{k_1 k_2}} \tag{41c}$$

As is seen from (41c), the waves $\hat{\alpha}_+(k,t), \hat{\alpha}_-(k,t)$ can be damped or amplified depending on the sign of $\chi^\pm$, i.e., on the direction of energy exchange between the two waves.

Furthermore, for simplification of the obtained results we assume that Fano asymmetry parameters $q_\nu$ and $q_\nu^L$ differ from each other slightly.

$$q_\nu^L = q_\nu(1 + \zeta_\nu), \tag{42}$$

where

$$|\zeta_\nu| = \left|\frac{\Delta q_\nu}{q_\nu}\right| \ll 1, \quad (\Delta q_\nu = q_\nu^L - q_\nu), \tag{43}$$

and

$$\frac{\left|\nu + i\frac{\gamma_c}{2}\right|}{|\tilde{\Omega}|} \ll 1. \tag{44}$$

If the conditions (43) and (44) are fulfilled, we obtain

$$\beta_1^L(\omega) = \beta_1(\omega) + b(\omega) \tag{45}$$

$$\beta_2(\omega) = \beta_1(\omega) + \frac{1}{2}b(\omega), \tag{46}$$

where

$$b(\omega) = -i\pi \sum_\nu \frac{1}{\mathrm{Im}S_\nu} \frac{\prod_k(s_\nu^* - \tilde{E}_k) + \frac{1}{2}\sum_j q_j \tilde{\Gamma}_j \prod_{k\neq\nu}(s_\nu^* - \tilde{E}_k)}{[\mathrm{Res}_\nu - \varepsilon_2 - \omega - i(\frac{\bar{\gamma}_2}{2} + \mathrm{Ims}_\nu)]\prod_{k\neq\nu}(s_\nu^* - s_k)(s_\nu^* - s_k^*)} \sum_j q_j \zeta_j \tilde{\Gamma}_j (s_\nu^* - \tilde{E}_k) \tag{47}$$

For storage of memory for photons it is necessary, that

$$N\tilde{g}^2 \gg |\tilde{\Omega}|^2. \tag{48}$$

Then we will arrive at

$$\omega_+(0) \approx 0 \tag{49a}$$

$$\omega_-(0) = N\tilde{g}^2\left(\mathrm{Im}\beta_1(\omega) + \frac{|\tilde{\Omega}|^2}{N\tilde{g}^2}\mathrm{Im}b(\omega) - \frac{\nu}{N\tilde{g}^2}\right) - iN\tilde{g}^2\left(\mathrm{Re}\beta_1(\omega) + \frac{\gamma_c/2}{N\tilde{g}^2}\frac{|\tilde{\Omega}|^2}{N\tilde{g}^2}\mathrm{Re}b(\omega)\right) \tag{49b}$$

$$v_g^{(+)} = c\frac{|\widetilde{\Omega}|^2}{N\widetilde{g}^2}\left[1 + \text{Re}f(\omega) + \frac{\gamma_c/2}{|\widetilde{\Omega}|^2}\frac{\text{Re}\beta_1(\omega)}{|\beta_1(\omega)|^2} - \frac{\nu}{|\widetilde{\Omega}|^2}\frac{\text{Im}\beta_1(\omega)}{|\beta_1(\omega)|^2}\right] \quad (49c)$$

$$\chi_+ = c\frac{|\widetilde{\Omega}|^2}{N\widetilde{g}^2}\left[\text{Im}f(\omega) - \frac{\gamma_c/2}{|\widetilde{\Omega}|^2}\frac{\text{Im}\beta_1(\omega)}{|\beta_1(x\omega)|^2} - \frac{\nu}{|\widetilde{\Omega}|^2}\frac{\text{Re}\beta_1(\omega)}{|\beta_1(\omega)|^2}\right] \quad (49d)$$

$$v_g^{(-)} = c\left[1 - \frac{|\widetilde{\Omega}|^2}{N\widetilde{g}^2}\text{Re}f(\omega) - \frac{\gamma_c/2}{N\widetilde{g}^2}\frac{\text{Re}\beta_1(\omega)}{|\beta_1(\omega)|^2} + \frac{\nu}{|\widetilde{\Omega}|^2}\frac{\text{Im}\beta_1(\omega)}{|\beta_1(\omega)|^2}\right] \quad (49e)$$

$$\chi_- = -\chi_+, \quad (49f)$$

where

$$f(\omega) = \frac{b(\omega)\beta_1^*(\omega)}{|\beta_1(\omega)|^2}. \quad (50)$$

It is seem from expressions (34a,b) and (36a,b), and also (49a-f) that $\hat{\alpha}_+(k, t)$ and $\hat{\sigma}_+(k, t)$ are the slow wave components of the quantum field and the spin wave, respectively, while $\hat{\alpha}_-(k, t)$- and $\hat{\sigma}_-(k, t)$-components propagate at high speed. It is necessary to notice, proceeding from expressions (49b) and (49f), that the fast components $\hat{\alpha}_-(k, t)$ and $\hat{\sigma}_-(k, t)$ decay rapidly in the medium and do not contribute to the considered process. Therefore, we will limit ourselves to only the slow components $\hat{\alpha}_+(k, t)$ and $\hat{\sigma}_+(k, t)$.

Consider the possibility of recording and reconstructing a quantum pulse in the case when the control field with interaction $\Omega$ is instantaneously switched off at $t < 0$ (Fig. 2). Prior to the moment of switching-off, the quantum pulse $\hat{\alpha}_+(k, t)$ emerges from the medium and coherence between the lower atomic states $\hat{\sigma}_+(k, t)$ is generated.

Under the EIT condition when

$$|\widetilde{\Omega}|^2|\beta_1(\omega)|T \gg 1, \quad (51)$$

the pulse passes through the transparency window without significant loss prior to the control-field being switched off ($f(\omega), \gamma_c$ and $\nu$ are very small) [1, 2]. Then, if condition (51) is satisfied, we obtain the following expression for the coherence between the lower states of the system using Eq. (31):

$$\hat{\sigma}_{21}(k,t) = -\left(1 - \frac{1}{2}f(\omega) - \frac{\gamma_c/2 - i\nu}{|\widetilde{\Omega}(t)|^2\beta_1(\omega)}\right)\frac{\widetilde{g}\hat{\alpha}(k,t)}{\widetilde{\Omega}(t)} + \frac{\mathscr{F}_{2,1}(k,t)}{|\widetilde{\Omega}(t)|^2\beta_1(\omega)}\left(1 - f(\omega) - \frac{\gamma_c/2 - i\nu}{|\widetilde{\Omega}(t)|^2\beta_1(\omega)}\right) \quad (52)$$

This coherence is maintained in the medium after the control field is switched off.

If the control field is switched on at $t = 0$ (Fig. 2), for the slow wave component $\hat{\alpha}_+(k, t)$ with the initial conditions

$$\hat{\sigma}_{21}(k,0) = \hat{\sigma}_{21}^{(0)}(k) = -\left(1 - \frac{1}{2}f(\omega) - \frac{\gamma_c/2 - i\nu}{|\widetilde{\Omega}(t)|^2\beta_1(\omega)}\right)\frac{\widetilde{g}\hat{\alpha}^{(0)}(k)}{\widetilde{\Omega}(t)} \quad (53)$$

$$\hat{\alpha}(k, 0) = 0 \quad (54)$$

we obtain the following expression:

$$\hat{\alpha}_+(k, t) = \left(1 - \frac{|\widetilde{\Omega}(t)|^2}{N\widetilde{g}^2}f(\omega) + \frac{i}{\beta_1(\omega)}\frac{\nu + \frac{i\gamma_c}{2}}{|\widetilde{\Omega}(t)|^2}\right)\hat{\alpha}^{(0)}(k)e^{k\chi_+ t}e^{-ikv_g^{(+)}t} + \hat{F}_f^{(+)}(k,t) \quad (55)$$

where $\hat{\alpha}^{(0)}(k)$ and $\hat{\sigma}_{21}^{(0)}(k)$ are the operator of the quantum pulse and the atomic operator, respectively, at the moment the control field is switched off. In initial condition (53) the Langevin noise operator equals zero at the initial moment due to condition (54) ($\hat{F}_f^{(+)}(k,t)$ is δ-correlated quantum operators of the Langevin noise for quantum field [21,22]).

Expression for the quantum field amplitude $\hat{\alpha}_+(k, t)$ in (55) shows that if $\chi_+ < 0$, the field damps and if $\chi_+ > 0$, it, on the contrary, is amplified at the expense of the energy received from fast components $\hat{\alpha}_-(k, t)$, as appears from (49f). From expression (55) follows that the initial pulse is distorted at the exit. Reconstruction of the initial quantum pulse is possible if the length of absorption (amplification) $l \sim \frac{1}{k\frac{\chi_+}{v_g^{(+)}}}$ exceeds the effective length $L \sim \frac{1}{k}$, or if $(l \gg L)$ $\frac{\chi_+}{v_g^{(+)}} \ll 1$.

It is seen from expression (55) that under this condition the shape and quantum state of the pulse are reconstructed after the coherent control field is switched on but the amplitude is lower than the original one:

$$\hat{a}_+(z,t) = \left(1 - \frac{|\tilde{\Omega}(t)|^2}{N\tilde{g}^2}f(\omega) + \frac{i}{\beta_1(\omega)}\frac{\nu + \frac{i\gamma_c}{2}}{|\tilde{\Omega}(t)|^2}\right)\hat{a}^{(0)}\left(z - v_g^{(+)}t\right) + \hat{F}_f^{(+)}(z,t) \quad (56)$$

## 4. THE CASE OF TWO EXCITON RESONANCES.

Now let us consider the case of two close lying isolated exciton resonances when:

$$\frac{\tilde{\Gamma}_k}{\tilde{E}_2 - \tilde{E}_1} \equiv p_k \ll 1, \quad k = 1,2 \quad (57)$$

and

$$\tilde{\Gamma}_1 = \frac{(\tilde{E}_1 - E_2 - \Delta_2)^2}{(\tilde{E}_1 - E_2 - \Delta_2)^2 + |\Delta_{12}|^2}\Gamma_1\left(1 + \frac{\Delta_{12}}{\tilde{E}_1 - E_2 - \Delta_2}\sqrt{\frac{\Gamma_2}{\Gamma_1}}\right)^2 \quad (58a)$$

$$\tilde{\Gamma}_2 = \frac{(\tilde{E}_2 - E_1 - \Delta_1)^2}{(\tilde{E}_2 - E_1 - \Delta_1)^2 + |\Delta_{12}|^2}\Gamma_2\left(1 + \frac{\Delta_{12}}{\tilde{E}_2 - E_1 - \Delta_1}\sqrt{\frac{\Gamma_1}{\Gamma_2}}\right)^2 \quad (58b)$$

$$\tilde{E}_{1,2} = \frac{1}{2}\left[E_1 + \Delta_1 + E_2 + \Delta_2 \mp \sqrt{(E_2 + \Delta_2 - E_1 - \Delta_1)^2 + |\Delta_{12}|^2}\right] \quad (58c)$$

$$\Gamma_k = 2\pi|U_{k,\lambda}|^2, \quad (k = 1,2) \quad (58d)$$

$$\Delta_{12}(\lambda) = P\int\frac{U_{1,E}U_{2,E}^*}{\lambda - E}dE \quad (58e)$$

Under the condition (57) the functions $\beta_1(x)$ and $b(x)$, where $x = \frac{\omega - (\tilde{E}_1 - \varepsilon_2)}{\Delta \tilde{E}}$ $(\Delta \tilde{E} = \tilde{E}_2 - \tilde{E}_1)$, have the following forms:

$$\beta_1(x) = \pi\left[1 - i\frac{p_1}{2}\frac{(1+iq_1)^2}{x+ip_1\left(x+\frac{1}{2}\right)} - i\frac{p_2}{2}\frac{(1+iq_2)^2}{x-1-ip_2\left(x-\frac{3}{2}\right)}\right] \quad (59a)$$

$$b(x) = \pi\left[p_1\frac{q_1(1+iq_1)}{x+ip_1\left(x+\frac{1}{2}\right)}\zeta_1 + p_2\frac{q_2(1+iq_2)}{x-1-ip_2\left(x-\frac{3}{2}\right)}\zeta_2\right] \quad (59b)$$

Fig. 3 graphically demonstrates the functions $\beta_1(x)$ and $b(x)$ in expressions (59a) and (59b) for various values of the Fano parameter q and for small parameters $p_1$, $p_2$, $\zeta_1$ and $\zeta_2$ in the vicinity of one-photon resonance. It is seen from the expressions (49c,d) and from the plots that for the functions $\beta_1(x)$ and $b(x)$ such that $\frac{\chi_+}{v_g^{(+)}} \sim 10^{-2}$, i.e., when the linear absorption (amplification) length of the pulse is much larger than the effective length, there is no distortion and the shape and quantum state of the light pulse are restored.

## 5. CONCLUSION

Quantum memory for photons and storage and reconstruction of quantum information is demonstrated for a crystal with impurity $\Lambda$-atoms in the case when the upper excited states are located in the exciton band. The configuration interaction of the upper states with the exciton band leads to the formation of a Fano-type resonant state similar to the autoionization atomic state. It is demonstrated that the quantum pulse propagates in the medium at a group velocity that is significantly less than the velocity of light in vacuum. We study the possibility for recording and reconstructing the quantum pulse upon the instantaneous on–off switching of the control field. In the case of the propagation and reconstruction of the quantum light pulse, its shape and quantum state remain unchanged, when the linear absorption (amplification) length is much longer than the effective length.

The authors acknowledge valuable discussion with G. G. Grigoryan and V. O. Chaltykyan.
This work was supported by the INTAS No 06-1000017-9234, ANSEF PS-opt-1347 and in part by the Scientific Research Foundation of the Government of the Republic of Armenia (Project No 96).

FIGURE CAPTIONS

Fig 1. Energy levels of the process under study.
Fig 2. Storage and reconstruction of the quantum pulse.
Fig 3. Plots of functions $\beta_1(x)$ and $b(x)$ in the case of two exciton resonances vs. dimensionless $x=\frac{\omega-(\widetilde{E}_1-\varepsilon_2)}{\Delta\widetilde{E}}$ in the case of $p_1=p_2=0.2$, $\zeta_1=\zeta_2=0.1$, for $q_1=7$, $q_2=4$ and $q_1=8$, $q_2=6$.

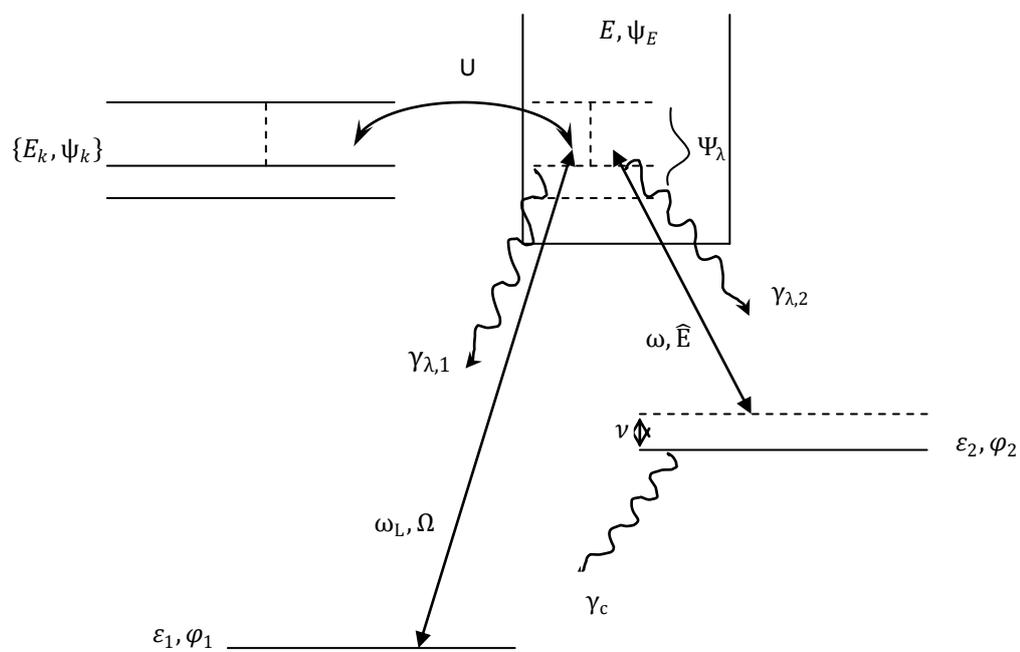

Fig 1

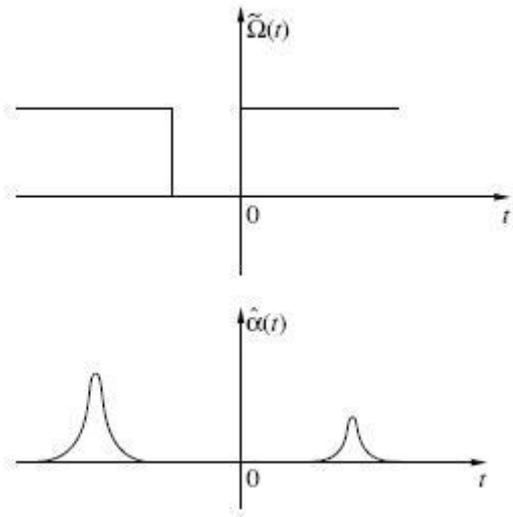

Fig 2

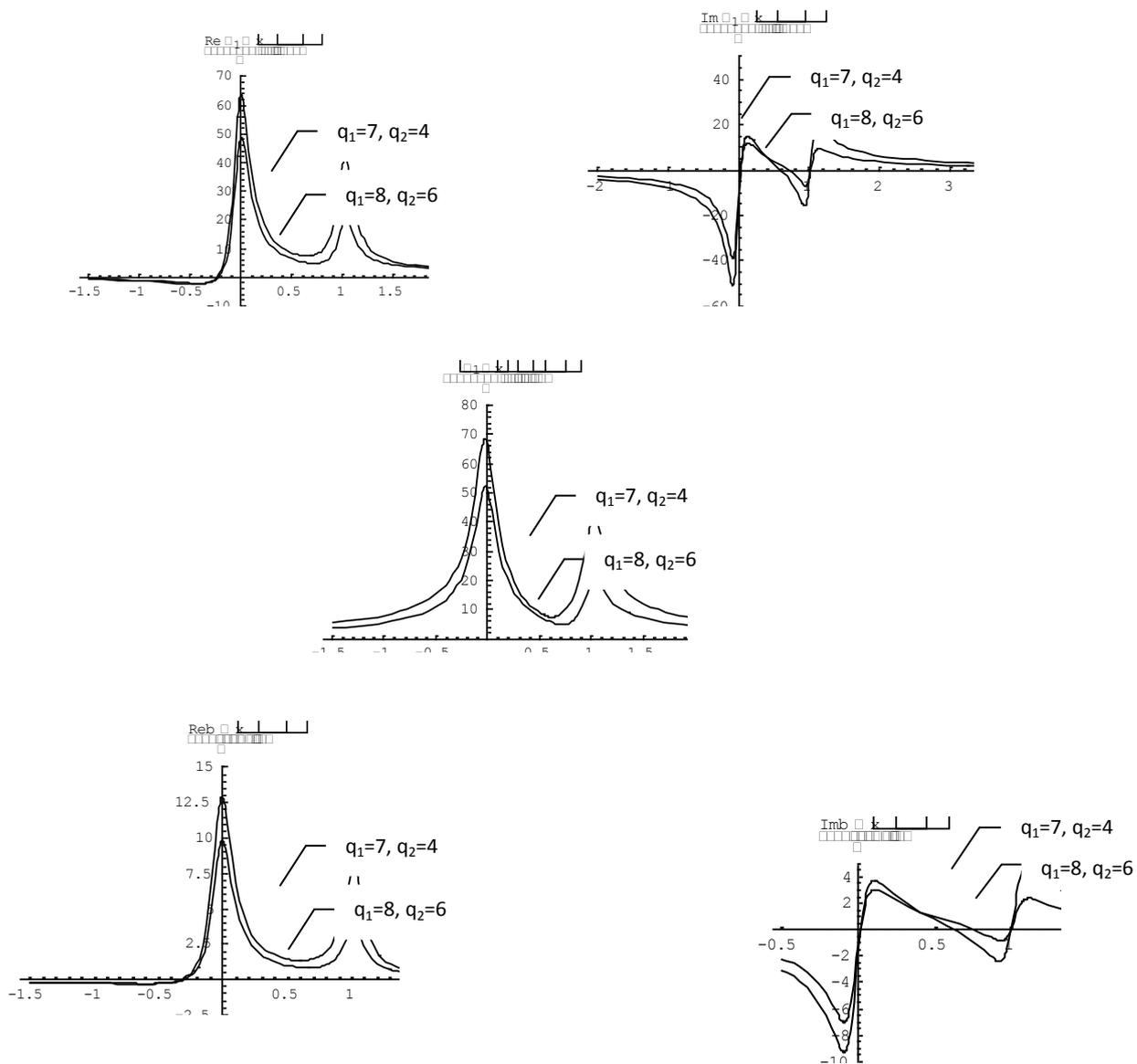

Fig 3